\documentclass[sigconf]{acmart}
\acmConference[ESEC/FSE 2021]{The 29th ACM Joint European Software Engineering Conference and Symposium on the Foundations of Software Engineering}{23 - 27 August, 2021}{Athens, Greece}
\usepackage{color}
\usepackage{transparent}
\usepackage{graphicx}
\usepackage{import}

\usepackage{framed}
\usepackage{color}
\usepackage{listings}
\usepackage{multirow}
\usepackage{comment}
\usepackage{fancyvrb}
\usepackage{xcolor}
\usepackage{balance}
\newboolean{showcomments}
\setboolean{showcomments}{true}
\ifthenelse{\boolean{showcomments}}
{ }
\ifthenelse{\boolean{showcomments}}
{ }
\ifthenelse{\boolean{showcomments}}
{ }
\AtBeginDocument{%
  \providecommand\BibTeX{{%
    \normalfont B\kern-0.5em{\scshape i\kern-0.25em b}\kern-0.8em\TeX}}}

\usepackage{tcolorbox}

\setcopyright{acmcopyright}
\copyrightyear{2021}
\acmYear{2021}
\acmDOI{10.1145/1122445.1122456}

\acmBooktitle{Woodstock '18: ACM Symposium on Neural Gaze Detection,
  June 03--05, 2018, Woodstock, NY}
\acmPrice{15.00}
\acmISBN{978-1-4503-XXXX-X/18/06}

\begin{document}
\def \toolname {\textit{BiasRV} }
\title[BiasRV]{BiasRV: Uncovering Biased Sentiment Predictions at Runtime}

\author{Zhou Yang, Muhammad Hilmi Asyrofi and David Lo}

\affiliation{%
  \institution{School of Computing and Information Systems}
  \country{Singapore Management University}
}
\email{{zyang, mhilmia, davidlo}@smu.edu.sg}

% \author{Muhammad Hilmi Asyrofi}
% \email{mhilmia@smu.edu.sg}
% \affiliation{%
%   \institution{Singapore Management University}
%   \country{Singapore}
% }

% \author{David Lo}
% \email{davidlo@smu.edu.sg}
% \affiliation{%
%   \institution{Singapore Management University}
%   \country{Singapore}
% }

\renewcommand{\shortauthors}{Yang et al.}

\begin{abstract}
  Sentiment analysis (SA) systems, though widely applied in many domains, have been demonstrated to produce biased results. Some research works have been done in automatically generating test cases to reveal unfairness in SA systems, but the community still lacks tools that can monitor and uncover biased predictions at runtime. This paper fills this gap by proposing \textit{BiasRV}, the first tool to raise an alarm when a deployed SA system makes a biased prediction on a given input text. To implement this feature, \toolname dynamically extracts a template from an input text and from the template generates gender-discriminatory mutants (semantically-equivalent texts that only differ in gender information). Based on popular metrics used to evaluate the {\em overall} fairness of an SA system, we define {\em distributional fairness} property for an {\em individual} prediction of an SA system. This property specifies a requirement that for one piece of text, mutants from different gender classes should be treated similarly as a whole. Verifying the distributional fairness property causes much overhead to the running system. To run more efficiently, \toolname adopts a two-step heuristic: (1) sampling several mutants from each gender and checking if the system predicts them as of the same sentiment, (2) checking distributional fairness only when sampled mutants have conflicting results. Experiments show that compared to directly checking the distributional fairness property for each input text, our two-step heuristic can decrease overhead used for analyzing mutants by $73.81\%$ while only resulting in $6.7\%$ of biased predictions being missed.
  Besides, \toolname can be used conveniently without knowing the implementation of SA systems. Future researchers can easily extend \toolname to detect more types of bias, e.g. race and occupation. The demo video for \toolname can be viewed at https://youtu.be/WPe4Ml77d3U and the source code can be found at https://github.com/soarsmu/BiasRV.
\end{abstract}




\maketitle

\section{Introduction}

Sentiment analysis (SA) systems \cite{MEDHAT20141093}, which aim to predict the sentiment of a given text, have been widely applied in many domains, e.g. predicting politics \cite{election} and healthcare \cite{NLPHealthcare}. However, evidence has shown that SA systems can be unfair and have gender bias. For example, Ribeiro et al. \cite{CheckList} found that an SA system fine-tuned on BERT always predicts negative when sensitive contents of a text template are filled with black, atheist, gay, and lesbian while predicting positive for Asian, straight, etc.
We use an example to demonstrate such discrimination. The following paragraph is a positive movie review from IMDb.

\begin{tcolorbox}
  \textit{"Dee Snider was inspired to do a two part song by a horror movie. This movie he wrote/directed/produced and starred in details the subjects from those songs (Horror-terria,from TwistedSister/ Stay Hungry). ...  This movie is perfect if you want something to give you nightmares and make you cringe about the possible and probable. IT COULD HAPPEN!!"}
  \end{tcolorbox}

An SA model fine-tuned on BERT \cite{BERT} predicts the sentiment of this paragraph as positive. However, if we generate a gender-discriminatory mutant by changing the name `Dee Snider' at the begining of this paragraph to `Lilly', which is usually used as a female name, the predicted result by the same model becomes negative. Such a case is not an exception: we change `Dee Snider' to 30 male names, e.g. Benedetto, and the results are all positive. We also replace the name with 30 female names, e.g. Julissa, but the predicted results are all negative. This provides a concrete example of gender bias, and if such bias happen, we want flag it.

Angell et al. \cite{Themis} believed that software fairness is part of software quality. To ensure the quality of SA systems, researchers propose some testing methods to uncover unfairness in NLP and SA systems, e.g. CheckList \cite{CheckList}, ECC \cite{EEC}, ASTRAEA \cite{soremekun2021astraea} and BiasFinder \cite{BiasFinder}. There is a simple metamorphic relationship behind these tools: modifying only sensitive contents of a text should not change predicted sentiment results. For example, in the movie review above, replacing `Dee Snider' with `Julissa' should make no difference.
These works, except BiasFinder \cite{BiasFinder}, all use pre-defined templates to generate texts with minor differences (we call them mutants). One template from CheckList \cite{CheckList} is  `{\verb|I am a {PROTECTED} {NOUN}|}', where the `{\verb|{PROTECTED}|}' placeholder can be replaced with black, white, gay, etc. and the `{\verb|NOUN|}' placeholder can be replaced with student, nurse, etc. These generated mutants can be used to test systems before deployment. The deployed SA systems receive many queries that normally mismatch pre-defined templates, making it challenging to detect biased predictions at runtime. Runtime verification (RV) is the process of checking whether each run of a system satisfies a given property \cite{VyPR2}. To the best of our knowledge, there is no tool that can uncover biased prediction of an SA system at runtime. 

In this paper, we propose \toolname to fill this gap. \toolname utilizes a mutation generation engine of BiasFinder \cite{BiasFinder} that can dynamically extract templates from input texts rather than rely on several pre-defined templates. As a result, \toolname can generate gender-discriminatory mutants (i.e., semantically-equivalent pieces of text that differ only in gender information) for queries received at runtime.\footnote{We focus on binary gender (male and female) and binary sentiment (positive and negative) but \toolname can be extended to non-binary scenarios too.} In the NLP community, researchers have proposed an evaluation metric to measure the {\em overall} fairness of an SA system~\cite{CheckList,EEC, huang-etal-2020-reducing}; the SA system is tested against a fixed and predefined set of gender-discriminatory mutants, and the distributions of sentiments predicted for male and female mutants are compared. However, this metric cannot be used to detect if an SA is biased towards a {\em specific} input text.  We tailor this evaluation metric and propose the distributional fairness concept specifying the requirement for a fair prediction that an SA system makes for an input text. For a piece of input text, distributional fairness requires that two sets of mutants (of the input text) from different gender classes to receive similar sentiment predictions. For example, it is acceptable that an SA system predicts $70\%$ males mutants as positive and $71\%$ female mutants as positive; while $70\%$ and $50\%$ are not acceptable since the difference between proportions of positive predictions exceeds a threshold, e.g. $10\%$.

Though distributional fairness can appropriately specify a fair prediction, it takes much time to verify. To reduce overhead, \toolname adopts a two-step heuristic: (1) sampling several mutants from each gender and check if the system predicts them as the same sentiment, (2) checking distributional fairness only when sampled mutants have conflicting results. The intuition is that if an SA system is biased towards {\em an input text}, many mutants shall be predicted as the opposite sentiments. When we sample these mutants, it is very likely to find conflicts and proceed to step (2) for more accurate but time-consuming verification. In contrast, if all the sampled mutants have the same result, there is only little chance that it is a biased prediction. Our evaluation results show that compared to directly checking the distributional fairness property for each input text, the 2-step heuristic can reduce overhead by $73.21\%$, while only causing $6.7\%$ of biased predictions to be missed. Low overhead is important especially for popular SA systems that are offered as a service (e.g., through a web API).

The rest of this paper is organized as follows. Section 2 describes the basic idea of the mutation generation engine in BiasFinder. Section 3 discusses the distributional fairness property and how \toolname is designed and used. Section 4 shows the evaluation results of \toolname on an SA system. In Section 5, we discuss some related work. Section 6 states some limitations of our tool. Finally, we conclude the paper and present future work in Section 5.
\section{Mutant Generation}
\label{biasfinder}

Our tool utilizes BiasFinder \cite{BiasFinder} to generate gender-discriminatory mutants. In this section, we briefly introduce the basic idea of how BiasFinder generates mutants.

Compared to previous works \cite{CheckList, EEC, soremekun2021astraea} that only generate test cases from limited numbers of handcrafted templates, BiasFinder can create templates dynamically from texts. This step is done by the {\em template generation engine} of BiasFinder.
Given a piece of text $I$ that can be viewed as a sequence of tokens $(t_1, t_2, \cdots, t_n)$, the template generation engine employs several NLP techniques, such as named entity recognition and coreference resolution, to identify the {\em protected tokens} $P(I)$. Protected tokens are tokens that divide a population into groups, e.g. gender, race or occupation. In this paper, we mainly discuss gender bias and limit protected tokens to names and gender pronouns \footnote{It is possible that no protected token is extracted. For example, no token in `I am happy' can reflect gender information.}. In Figure \ref{fig:genderbiasfinder}, $\langle${\verb|Drew Barrymore|}, {\verb|she|}, {\verb|her|}$\rangle$ are identified protected tokens. Then BiasFinder substitutes $P(I)$ with placeholders, just like the generated template in Figure \ref{fig:genderbiasfinder}.

Another engine in BiasFinder is called {\em mutant generation engine} that generates gender-related mutants by replacing placeholders, i.e. $P(I)$ with gender-specific tokens. For example, in Figure \ref{fig:genderbiasfinder}, we use $\langle${\verb|James|}, {\verb|he|}, {\verb|his|}$\rangle$ to generate a male mutant and use $\langle${\verb|Anne|}, {\verb|she|}, {\verb|her|}$\rangle$ to generate a female mutant. It should be noted that all the tokens in $P(I)$ should be changed correspondingly, which means $\langle${\verb|Anne|}, {\verb|he|}, {\verb|her|}$\rangle$ is invalid modification because the subjective personal pronoun `{\verb|he|}' conflicts with the objective personal pronoun `{\verb|her|}'. BiasFinder also ﬁlters the names to make sure that the selected names are only used for one gender globally. In the default setting, BiasFinder will generate 30 mutants for each gender if an extracted $P(I)$ contains protected tokens.

\begin{figure}[t]
  {\vspace{-0.5em}}
  \begin{framed}
  \begin{flushleft}
  \small
  \noindent{\textbf{Text}}
  \newline\noindent{'Never Been Kissed' is a real feel good film. \underline{\textbf{Drew Barrymore}} is excellent again, \underline{\textbf{she}} plays \underline{\textbf{her}} part well. I felt I could relate to this film because of the school days I had were just as bad. 
  }
  \vspace{0.1cm}\newline\noindent{\textbf{Generated Template}}
  \newline\noindent{'Never Been Kissed' is a real feel good film. \underline{$\langle \mathit{\textbf{name}}\rangle$} is excellent again, \underline{$\langle \mathit{\textbf{subjective}}$-$\mathit{\textbf{pronoun}}\rangle$} plays \underline{$\langle \mathit{\textbf{possesive}}$-$\mathit{\textbf{pronoun}}\rangle$} part well. I felt I could relate to this film because of the school days I had were just as bad.
  }
  \vspace{0.1cm}\newline\noindent{\textbf{Male Mutant}}
  \newline\noindent{'Never Been Kissed' is a real feel good film. \underline{\textbf{James}} is excellent again, \underline{\textbf{he}} plays \underline{\textbf{his}} part well. I felt I could relate to this film because of the school days I had were just as bad.
  }
  \vspace{0.1cm}\newline\noindent{\textbf{Female Mutant}}
  \newline\noindent{'Never Been Kissed' is a real feel good film. \underline{\textbf{Anne}} is excellent again, \underline{\textbf{she}} plays \underline{\textbf{her}} part well. I felt I could relate to this film because of the school days I had were just as bad.
  }
  \end{flushleft}
  \end{framed}
  {\vspace{-1em}}
  \caption{An illustrative example of how BiasFinder generate bias-discriminatory mutants.}
  \label{fig:genderbiasfinder}
\end{figure}

\section{B\lowercase{ias}RV}

\toolname is a tool that can uncover potentially biased predictions that an SA system makes at runtime. Like other runtime verification tools, in this section, we need to specify the property that an unbiased prediction should satisfy. Then we discuss how \toolname checks the property more efficiently, with a minimal trade off of a small number of biased predictions being missed. It should be noted that for simplicity and consistency, we mainly use gender bias as examples in the paper. But \toolname can be extended to uncover other types of discrimination, e.g. race bias and occupation bias.

\subsection{{Distributional Fairness}}
We define the {\em distributional fairness} concept for an SA system. Distributional fairness is described as the goal that for one piece of text, mutants from different gender classes should be treated similarly as a whole. Here `similar treatment ' refers to the expectation that the distribution of predicted sentiments for the two groups of mutants should be close. For example, it is acceptable that for a given input text, an SA system predicts $70\%$ male mutants as positive and $71\%$ female mutants as positive. However, we think it is unfair if $70\%$ males mutants are predicted as positive while only $50\%$ female mutants are predicted as positive. The difference between proportions of positive prediction exceeds a threshold, e.g. $10\%$.

% 也许可以直接扔掉group fairness
% In the NLP community, researchers have proposed an evaluation metric to measure the {\em overall} fairness of an SA system; the SA system is tested against a fixed and predefined set of gender-discriminatory mutants, and the distributions of sentiments predicted for male and female mutants are compared \cite{CheckList,EEC, huang-etal-2020-reducing}. However, this metric cannot be used to detect if an SA is biased towards a {\em specific} input text.
% compare the average predicted results, compare the average predicted results of texts 
Previous works in NLP~\cite{CheckList,EEC, huang-etal-2020-reducing} evaluate the {\em overall} fairness of an SA system using a {\em fixed and predefined set of mutants}. Specifically, the distributions of sentiments predicted for the male and female mutants are compared and a large difference (in the distributions) corresponds to a biased SA system. This evaluation metric is also similar to group fairness concept proposed by previous researchers \cite{fairway, two_fairness}, which is described as the goal that privileged and unprivileged groups are treated similarly. It can be used to measure an algorithm's overall fairness, but cannot decide whether the algorithm makes a biased prediction on {\em a specific input}. Distributional fairness, which is defined on the generated mutants of a specific input text, can specify the requirement whether  a fair prediction is made for that specific input text at runtime.

We illustrate the distributional fairness concept for SA systems with the following notations. Assuming we have a group of male mutants $M$ and a group of female mutants $F$ generated from an original input text $I$, we expect that for both genders the proportions of mutants predicted as positive should be close enough. Formally speaking, the following property should be satisfied:

\begin{equation}
  |pos_F - pos_M| \leq \alpha
\end{equation}

In the above formula, $pos_F$ is the proportion of female mutants predicted as positive (ranging from $0$ to $1$), and $pos_M$ can be similarly computed for male mutants. $\alpha$ is a threshold representing our tolerance of difference in SA systems' predictions on male and female mutants. Smaller $\alpha$ means less tolerance and our expectations for SA systems having more similar results for mutants of two genders. By default, we set the value of $\alpha$ as $0.10$. In practice, the value of $\alpha$ can be set based on the sensitivity of the target system being monitored.

\subsection{Uncover Bias}
\label{uncover}

\begin{figure}[t!]
	\centering
	\includegraphics[width=1\linewidth]{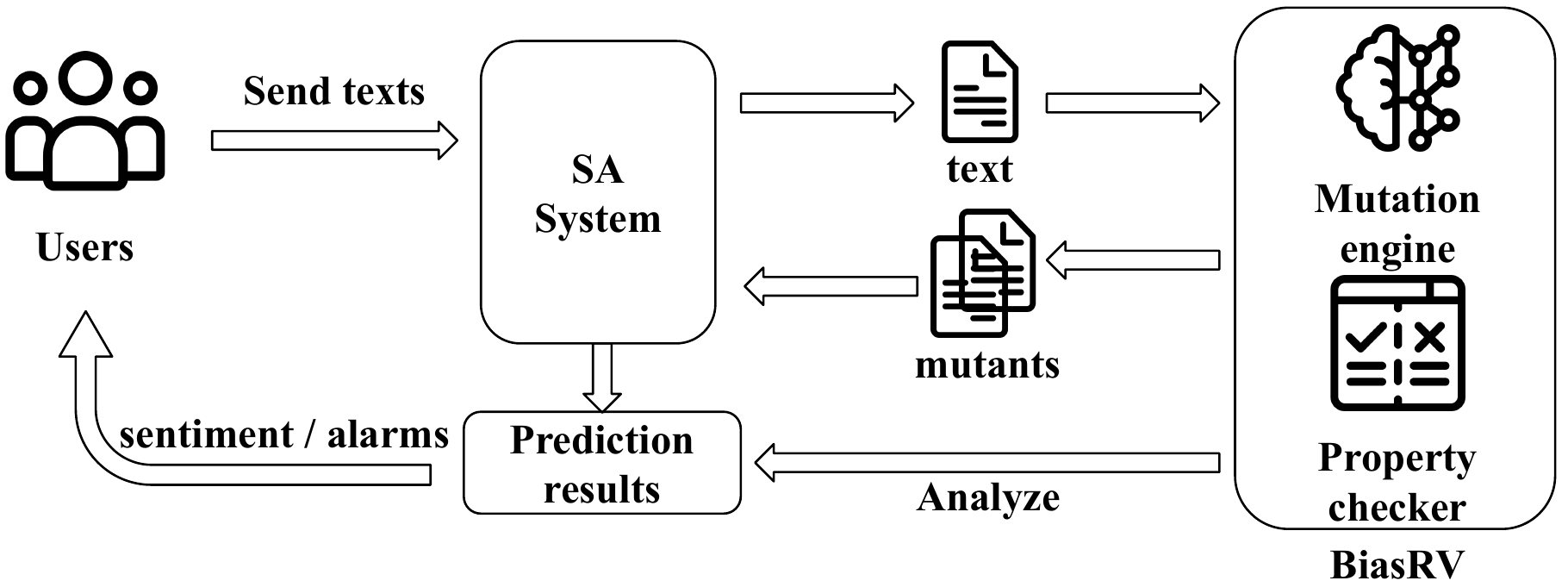}
	\caption{Monitoring an SA system with \textit{BiasRV}.}
	\label{fig:workflow}
\end{figure}

We introduce how \toolname uses the distributional fairness to monitor SA systems and uncover biased predictions. The overall workflow of \toolname is illustrated in Figure \ref{fig:workflow}. Users send text queries to a deployed SA system and expect the system to return the predicted sentiment of the query. Like other runtime verification tools, \toolname needs to collect some events of the running system. First, it fetches the text and returns generated mutants to the SA system to predict. Then \toolname analyzes whether the distributional fairness property is satisfied and raises alarms if a biased prediction is uncovered.

We expect a good runtime verification tool to be able to report violations to the checked property accurately and yet still incur have low overhead. As mentioned earlier, although we think distributional fairness can specify unbiased prediction requirement for an SA system, the major drawback is that it takes much time to analyze all mutants and compute distributional difference of each gender. So we propose a two-step heuristic to make \toolname uncover biased predictions more efficiently. We describe how the two-step heuristic works as follows.

In the first step, we randomly select no more than $X$ mutants from each gender and check whether these mutants are all predicted as of the same sentiment. If all the predicted sentiments are all the same, it is less likely to be a biased prediction. If at least one of them has a different predicted sentiment from the others, we proceed to the second step to check whether distributional fairness is satisfied. The intuition is that if an SA system makes a potentially biased prediction on an input text, it shall predict many mutants having different sentiments. When we have such a case, the likelihood of a biased prediction is higher. Then we can proceed to step 2 for more accurate but time-consuming verification. With the two-step heuristic, we can reduce overhead by filtering the cases that are less likely to be biased predictions. However, though the possibility is relatively low, \toolname might miss reporting biased cases. Users can make a trade-off between overhead and accuracy by adjusting the value of $X$. Larger $X$ can lead to more accurate results but introduce more overhead. By default, we set the value of $X$ for \toolname as $4$.

\subsection{A Use Example}
Users can use a simple command `{\verb|pip install bias_rv|}' to install \toolname easily. To uncover biased predictions at runtime, we need to wrap the API for predicting sentiment with the `{\verb|verify()|}' function in {\verb|bias_rv|} package. The following code segment illustrates a simple usage case.

\begin{lstlisting}[language=Python, basicstyle=\ttfamily\small]
from bias_rv.BiasRV import biasRV

# sa_system.predict() takes a piece of text 
# and return its sentiment
rv = biasRV(sa_system.predict,X=4,alpha=0.10)

result, is_bias = rv.verify(text)
  \end{lstlisting}

First, we need to import \toolname and then instantiate it. To create a verifier ({\verb|rv|}), we need to pass a function as a parameter. This function {\verb|sa_system.predict()|} takes a piece of text as input and returns its predicted sentiment. Besides, we need to specify parameters discussed in Section \ref{uncover} (i.e. $X$ and $\alpha$). Then, we use {\verb|rv.verify()|}  to wrap the original  {\verb|predict()|} function.  {\verb|rv.verify()|} can return an additional value,{\verb|is_bias|}, to indicate whether a biased prediction happens. Instantiating a verifier requires no implementation details of {\verb|sa_system.predict()|}, so \toolname can be used by both the SA service provider at server end and users at the client end.

\section{Evaluation}

We apply \toolname to an SA system that is constructed by fine-tuning a pre-trained BERT model \cite{BERT}. The SA system can achieve 92.0\% accuracy on 25,000 pieces of IMDB movie reviews unseen during training. We analyze the performance of \toolname by investigating the following research questions:

\vspace{0.2cm}\noindent\textbf{RQ1. } {\em Can \toolname detect biased predictions at runtime?}

To address RQ1, we analyze all the sentiment predictions that \toolname labeled as potentially biased. We run an SA system and use \toolname to monitor the system. We send 25,000 different texts as queries to the SA system. The 25,000 queries come from the IMDB movie review test set that the SA system has not seen during training. We set the parameters of our 2-step heuristics ($X$ and $\alpha$) as $4$ and $0.10$ respectively. We find that \toolname can generate gender-discriminatory mutants for $3,042$ texts (the remainder of the 25,000 texts include no protected tokens).
 \toolname detects $15$ biased predictions out of the $3,042$ texts.

\vspace{0.2cm}\noindent\textbf{RQ2. } {\em How much overhead does \toolname incur? Can the 2-step heuristic lead to a lower overhead?}

When processing an input text, \toolname introduces two main time overhead: time to generate mutants and time to analyze mutants. In the 25,000 test queries, BiasFinder needs from $0.009$s to $8.01$s to generate mutants. The time required increases linearly with the length of input texts and is mainly caused by the coreference resolution step in \textit{BiasFinder}. Optimizing \textit{BiasFinder} is not the main focus of this paper, so we pay more attention to the other overhead. If we verify distributional fairness specification for all the input texts, it will introduce $6.838$ times overhead compared to analyzing the original text on average. The two-step heuristic first samples several mutants to check if the system predicts them as the same sentiment and verifies distributional fairness using all mutants only when sampled mutants have conflicting results. We measure the overhead caused by directly checking the distributional fairness property, and by employing our two-step heuristic. When we set $X$ as $4$ and $\alpha$ as $0.10$, the two-step heuristic can decrease the overhead by $73.81\%$ while only misses reporting $6.7\%$ of biased predictions.

\section{Related Work}
The closest work to ours is \textit{BiasFinder} \cite{BiasFinder}. It provides the mutation generation engine used in this paper. \textit{BiasFinder} aims at using metamorphic relationships to find failed test cases revealing that an SA system has a bias. Section 2 provides more detailed information about \textit{BiasFinder}.

We briefly introduce other work proposed to uncover discrimination in AI systems. Themis \cite{Themis}, Aeqitas \cite{Aeqitas}, FairTest \cite{FairTest} and Fairway \cite{fairway} aim at uncovering bias in software systems that take tabular data as input.
There are some papers and tools designed to reveal bias in NLP-related systems. CheckList \cite{CheckList} uses a limited number of pre-deﬁned templates to generate test cases and show that an SA system fine-tuned on BERT always predicts negative when the templates are filled with black, atheist, gay, and lesbian.
Kiritchenko and Mohammad \cite{EEC} presented Equity Evaluation Corpus (EEC), which consists of 8,640 English sentences generated from 11 templates. However, EEC is criticised for relying on pre-deﬁned templates that may be too simplistic \cite{9260964}.
A more recent tool is ASTRAEA \cite{soremekun2021astraea}, which leverages context-free grammar to generate discriminatory inputs that reveal fairness violations in software systems. ASTRAEA can generate more diverse texts, but essentially it still relies on pre-defined templates that must adhere to languages defined by the context-free grammar. 

% For more comprehensive literature review on mitigating gender-bias in software systems, please refer to the survey by Sun et al.~\cite{sun-etal-2019-mitigating}

To the best of our knowledge, there is no runtime verification tool that can uncover biased predictions made by an SA systems after deployment. The testing works mentioned above mainly use metamorphic relationships to discover biased predictions for known texts, i.e. texts generated from pre-defined templates. But such templates are static, and at runtime, SA systems can receive texts that mismatch these templates, which makes it challenging to build a runtime verification tool. \textit{BiasFinder} addresses this limitation and can dynamically generate templates for any given text. It is the foundation that \toolname uses to monitor fairness at runtime.

\section{Threats to Validity and Limitation}
The template generation engine used in \toolname employs named entity recognition and coreference resolution to identify protected tokens, which are still under active research. It may generate invalid mutants. Replacing names and genders can change semantics.
\section{Conclusion}
In this paper, we present \textit{BiasRV}, a tool that can uncover potentially biased predictions made by an SA system at runtime. \toolname can extract and replace gender information in a piece of text to generate gender-discriminatory mutants. Then it queries SA systems with these mutants to get predicted sentiments. We propose the distributional fairness property for specifying an unbiased prediction made by an SA system at runtime. But verifying the distributional fairness property can cause much overhead to the system. So \toolname adopts a two-step heuristic to uncover potentially biased predictions at a lower time cost and still maintain a low rate of miss reporting. We apply \toolname to an SA system. We find that it can find $xx$ biased predictions from 25,000 texts. Also, we find that our two-step heuristic is effective in reducing overhead by $73.81\%$, while only causing $6.7\%$ of biased predictions to be missed. We plan to support \toolname with more types of bias (e.g., race, occupation, etc.) and optimize BiasFinder's mutation generation engine to achieve an even lower overhead.

\begin{acks}
  This research was supported by the Singapore Ministry of Education (MOE) Academic Research Fund (AcRF) Tier 1 grant.
\end{acks}

\balance
\bibliographystyle{ACM-Reference-Format}
\bibliography{reference}

\appendix

\end{document}